\documentclass[a4paper,11pt]{article}
\usepackage{setspace}
\usepackage{authblk}

\usepackage{fullpage}
\usepackage{graphicx}
\usepackage{subfigure}
\usepackage{epsfig}
\usepackage{dcolumn}
\usepackage{bm}

\usepackage{amsmath,amssymb,cite}

\begin{document}
 
 \title{Second law for an autonomous information machine connected with multiple baths}
\author{Shubhashis Rana\footnote{email: shubhashis.rana@gmail.com}}
\affil{S. N. Bose National Centre for Basic Sciences, JD Block, Sector-III, Salt Lake City, Kolkata - 700 106, India}

\maketitle{}






\newcommand{\nwc}{\newcommand}
\nwc{\vs}{\vspace}
\nwc{\hs}{\hspace}
\nwc{\la}{\langle}
\nwc{\ra}{\rangle}
\nwc{\lw}{\linewidth}
\nwc{\nn}{\nonumber}

\nwc{\pd}[2]{\frac{\partial #1}{\partial #2}}
\nwc{\zprl}[3]{#3~Phys. Rev. Lett. ~{\bf #1},~#2}
\nwc{\zpre}[3]{#3~Phys. Rev. E ~{\bf #1},~#2}
\nwc{\zpra}[3]{#3~Phys. Rev. A ~{\bf #1},~#2}
\nwc{\zjsm}[3]{#3~J. Stat. Mech. ~{\bf #1},~#2}
\nwc{\zepjb}[3]{#3~Eur. Phys. J. B ~{\bf #1},~#2}
\nwc{\zrmp}[3]{#3~Rev. Mod. Phys. ~{\bf #1},~#2}
\nwc{\zepl}[3]{#3~Europhys. Lett. ~{\bf #1},~#2}
\nwc{\zjsp}[3]{#3~J. Stat. Phys. ~{\bf #1},~#2}
\nwc{\zptps}[3]{#3~Prog. Theor. Phys. Suppl. ~{\bf #1},~#2}
\nwc{\zpt}[3]{#3~Physics Today ~{\bf #1},~#2}
\nwc{\zap}[3]{#3~Adv. Phys. ~{\bf #1},~#2}
\nwc{\zjpcm}[3]{#3~J. Phys. Condens. Matter ~{\bf #1},~#2}
\nwc{\zjpa}[3]{#3~J. Phys. A: Math theor  ~{\bf #1},~#2}

\begin{abstract}

 In an Information machine  system's  dynamics  gets  affected  by the attached information reservoir.
 Second law  of thermodynamics can be apparently violated for this case. In this article  we have derived second law 
 for an information machine, when the  system   is connected to multiple heat baths along with a work source and
 a single information reservoir. Here a sequence of bits written on a tape is considered 
 as an information reservoir. We find that
 the bath entropy production during a time  interval is restricted by the change of Shannon entropy of the composite
 system (system +  information reservoir) during that interval. We have also given several examples where this
 law can be applicable and shown that our bound is tighter.
\end{abstract}

 \hspace{0.2cm}Keywords: information processing, exact results, stochastic processes

\section{Introduction}

Second law  is a fundamental law in thermodynamics and always valid on an average \cite{jar97,cro98,sai12}.
According to the law average entropy production is always positive. Recent 
development of fluctuation theorems dictates that it is possible to find negative entropy production 
for an individual event for any  duration although their probabilities are  exponentially small.
However  validity of the second law was questioned by Maxwell even more than a century ago, 
when he proposed a thought experiment involving an intelligent being known as Maxwell demon \cite{max71}.
In this gedanken experiment only by
knowing the velocity of gas particles confined in a box, a demon can separate them into hotter
(consists of faster molecules) and colder (consists of slower molecules) part without doing any work. 
Half a century later, Szilard proposed another thought experiment \cite {szi29} where he showed that it is always
possible to extract heat from single heat bath and perform useful work cyclically 
when a gas molecule, confined in a box, is treated by a certain protocol which involves measurement
of the state of the system. 

To understand these puzzles, the last century has witnessed  several wonderful research works 
establishing the connection between information theory and thermodynamics. 
\cite{lan61,ben82,zur89,leff03,ved09,man12,bar13,bar14,jar13,sag09,par15}. In fact one needs to take into account 
the cost of information during the process and  above all, the process would be completed only when 
the information  contained in the memory register of the demon will be erased. Now, according
to  Landauer\cite{lan61}, one need to do at least  $k_B T \ln 2$ work to erase one bit of information ($k_B$ represents
Boltzmann constant). Hence the second law is saved when one takes into account the effect of information.

There are mainly two cases in the framework of information processing when the second law is apparently
violated \cite{bar14,jar13}. In the first one, measurement is 
performed and depending on the measurement outcome, the protocol is altered. In another approach, 
the information contained in an information reservoir is  changed when it is allowed 
to interact sequentially with the system \cite{man12,man13,ran16}. A sequence of bits written on a tape 
can be considered as an information reservoir. Recently  the second type of approach
draws many attention and second law has been derived consequently\cite{man13,boyd16,boyd17,mer15,mer16}. The performance of 
autonomous information machine  has been explained just taking  bath entropy production
 restricted by the configuration entropy change of the tape\cite{bar13,bar14,man12,man13,ran16}. However the inequality should 
contain the correlation between demon and tape which makes the inequality stronger\cite{man13}. 
In \cite{boyd16,boyd17} it is shown that the work done is bounded by the change in Kolmogorov-Sinai dynamical entropy
rate of the tape when system is connected to a heat bath, a work source and an information reservoir(tape).
Here, the derivation is done by taking into account the correlations within the input string
and  those in the output string  generated during its evolution connecting with the demon. 
Note that, this bound is stronger  when input is uncorrelated or the system(ratchet or demon) is memoryless (i.e.,
it has no internal states), compared to the earlier version \cite{bar13,bar14,man12} where statistical correction 
between the bits are neglected and work is bounded by the marginal configuration Shannon entropy
of the individual bits. Although the result is generally valid even when input is correlated
and the ratchet has memory there is some concern as 
discussed in \cite{mer16}. Besides  the second law is derived  in  \cite{mer16} exactly in a 
straight forward  way by simply adding  the inequalities for each individual cycles. 
When  a system is connected to a heat bath, a work source and an information reservoir, 
the corresponding  second law for single cycle is given by

\begin{equation}
 \beta W \le \mathcal{H}(\Pi_{\tau}) - \mathcal{H}(\Pi_{0}).
 \label{mer-res}
\end{equation}
Here $\mathcal{H}$ represents the Shannon entropy of the joint system consisting of interacting bit and system. If
that joint distribution at any time t is given by $\Pi_{t}$ then corresponding Shannon entropy is
denoted by $\mathcal{H}(\Pi_t)=-\sum \Pi_t\ln \Pi_t$ where the sum is done over all possible states.
The above equation dictates that the average 
extracted work $W$ during the time interval $0\leq t<\tau$  is restricted by the change 
of Shannon entropy of the joint system during that interval. Motivated by this,  we would like to study
an information machine which is connected to multiple baths.  We have obtained corresponding second law 
which is exact and most general as well as the bound is more compact. The derivation here is done by scrutinizing 
how an autonomous information machine processed a tape sequentially during its operation. Moreover the obtained inequality 
reduced to the earlier results in spacial cases. The organization of the paper is as follows.
First we  describe the model and derive our result. Then we compare the result with the earlier studies.
After that we give several examples where this law can be applicable. Finally taking a simple model, we numerically
showed that our bound is  stronger. 

\section{The Model}

 The model consists of a demon(system) which is attached with an information reservoir and a work source.
 A tape, where the information is written as  discrete symbols, acts as an information reservoir.
 The input tape is formed by a sequence of symbols $x_1$,$x_2$,$x_3$,...,$x_N$ which is taken 
 from a finite set $\chi$ (for binary symbols $\chi=(0,1)$). Each symbol  interacts one by 
 one sequentially with the demon.  As a result, the demon state is going through internal states
 $s_1$, $s_2$, $s_3,...,s_N$ which is taken from a finite set $S$. On the other hand, 
 after interaction, the outgoing tape consists of another sequence of symbols $y_1$,$y_2$,$y_3$,...,$y_N$
 which are elements of same set $\chi$. Note that there are no intrinsic transitions in the tape. Only during its 
 interaction with the demon the tape state may change.
 The total entropy of the incoming tape is given by
 \begin{equation}
  H(X^N)=-\sum_{x^N\in \chi}P(x^N)\ln P(x^N).
 \end{equation}
 Here, $P(x^N)=P(x_1,x_2,x_3,...,x_N)$ is the probability distribution of that sequence of the input.
 Now, If the input sequence is correlated then 
 \begin{eqnarray}
  H(X^N)  &&=-\sum_{x^N\in \chi}P(x^N)\ln P(x^N)\nn\\
  &&=-\sum_{x^N\in \chi} P(x^N)\ln \left[P(x_N|x^{N-1})......P(x_3|x_2,x_1)P(x_2|x_1)P(x_1)\right]\nn\\
  &&=-\sum_{x^N\in \chi} P(x^N)\ln P(x_N|x^{N-1})-.....-\sum_{x^N\in \chi} P(x^N)\ln P(x_3|x_2,x_1)\nn\\
 && \hspace{3cm}-\sum_{x^N\in \chi} P(x^N) \ln P(x_2|x_1)-\sum_{x^N\in \chi} P(x^N)\ln P(x_1)\nn\\
 &&=-\sum_{x^N} P(x^N)\ln P(x_N|x^{N-1})-.....-\sum_{x_3,x_2,x_1} P(x_3,x_2,x_1)\ln P(x_3|x_2,x_1)\nn\\
 && \hspace{3cm}-\sum_{x_2,x_1} P(x_2,x_1) \ln P(x_2|x_1)-\sum_{x_1} P(x_1)\ln P(x_1)\nn\\
 &&=\sum_{n=1}^{N} H(X_n|X^{n-1}).
 \end{eqnarray}
 Where $H(X_n|X^{n-1})=-\sum_{x^n} P(x^n)\ln P(x_n|x^{n-1})$. Similarly if $P(y^N)=P(y_1,y_2,y_3,...,y_N)$
 represents the probability distribution of the output sequence of the tape, then its entropy 
 is given by
 \begin{equation}
  H(Y^N)=-\sum_{y^N\in \chi}P(y^N)\ln P(y^N)=\sum_{n=1}^{N} H(Y_n|Y^{n-1}).
 \end{equation}

 The demon can interact with the nearest symbol of the tape at a time. Now if the tape 
 moves with constant speed $v$ then each symbol gets $\tau$ time to interact, after that the next
 symbol arrives. During that time, the joint state of demon and tape evolves with time.
 As an example, in $n^{th}$ interaction interval in between time  $(n-1) \tau \leq t < n \tau$,
 the input joint state $(x_n,s_n)$ evolves and finally reaches to $(y_n,s_{n+1})$.
 In the beginning of the next cycle, the demon state does not change but the tape is advanced 
 by one unit. As a result the next cycle  starts from  the joint state $(x_{n+1},s_{n+1})$.
 After time $\tau$, the state  $(x_{n+1},s_{n+1})$  evolves to  $(y_{n+1},s_{n+2})$ and this process 
 continues until  the tape  passes completely.

\begin{figure}[!ht]
\vspace{0.5cm}
\begin{center}
 \includegraphics[width=4cm]{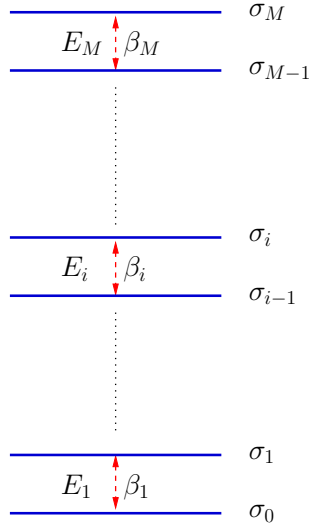}
\caption{Schematic diagram to describe the joint states of demon and information reservoir.
The energy difference between the states $\sigma_i$ and $\sigma_{i-1}$
 is denoted by $E_i$ and when transition occurs between these states, energy is exchanged only with the bath 
having inverse temperature $\beta_i$,}
\label{fig1}
\end{center}
\end{figure}

 Note that there is two types of dynamics. One is discrete  and only deals 
 with the input and output states ( $(x_1,s_1) \rightarrow (y_1,s_2);(x_2,s_2) \rightarrow (y_2,s_3);....$ ).
 Another one is continuous and deals how  a output state is evolved from the input state during
 the time interval $\tau$.
 Consider the states of the composite system of demon and interacting tape are taken  form a product set
  ($\chi \times S$)   that contains  M + 1 elements which are denoted by $(\sigma_0,\sigma_1,.....,\sigma_M)$. 
  Now at the starting of  $n^{th}$ cycle  the input 
 state is related to the joint state by $(x_n,s_n)\equiv \sigma^{(n-1)\tau}$.
 Note that the superscript only denotes the time at which the state appears.  
The state $\sigma^{(n-1)\tau}$ then evolves according to the dynamics and finally 
 at $n\tau ^{-}$ it reaches to another element, say $\sigma^{n\tau^{-}}\equiv (y_n,s_{n+1})$.
 At time $n\tau$ the tape is forwarded by one unit and the bit state is changed.
 As a result, the next cycle starts from 
 $(x_{n+1},s_{n+1})\equiv\sigma^{n\tau}$ which is again an element of that set 
 and the process continues. It can be mentioned that if the output and next input symbol 
 (bit for binary sequence) is same then the  actual state does not change
 at the time of this switching, if not, then it starts from  another 
 element of $\sigma$. Next  we will describe the dynamics and the evolution of the joint system
 in a particular interval in detail.

 \section{Derivation of Second Law in presence of multiple baths}
 
 The energy difference between the states $\sigma_0$ and $\sigma_1$ is $E_1$ (fig.\ref{fig1}). Similarly
 for  $\sigma_1$ and $\sigma_2$ it is $E_2$ and so on. Note that, we have taken $\sigma_0$ as
 ground state and corresponding energy is taken as 0.  Each of these consecutive states can exchange energy with only one bath and 
 transition can only happen between the consecutive states. As an example, transition between  $\sigma_0$
 and $\sigma_{1}$  can only happen by exchanging heat from a bath with inverse temperature $\beta_1$. 
 Hence corresponding transition rates satisfy  the required detailed balance \cite{van}
\begin{equation}
\frac{ R_{\sigma_0\rightarrow\sigma_1}}{ R_{\sigma_1\rightarrow\sigma_0}}=\exp^{-\beta_1 E_1}.
\end{equation}
  Similarly for $\sigma_{1}$  and $\sigma_{2}$ the transition can happen when heat is exchanged from the
 bath with inverse temperature $\beta_2$ and corresponding transition rates satisfy detailed balance
\begin{equation}
\frac{ R_{\sigma_1\rightarrow\sigma_2}}{ R_{\sigma_2\rightarrow\sigma_1}}=\exp^{-\beta_2 E_2},
\end{equation}
 and so on. Note that according to the construction of the model there is no other transition  possible from or to 
 $\sigma_1$. Moreover, the network of all the states form a linear chain and only back and forth excursions 
 (transitions) along the chain is possible.

During any time interval $\tau$, the probability distribution of joint states of demon and bits $\Pi_t(\sigma)$ 
evolves according to the master equation 
\begin{equation}
 \frac{d\Pi_t(\sigma)}{dt}=\mathcal{R}\Pi_t({\sigma}),
\end{equation}
where $\mathcal{R}$ denotes the transition rate matrix whose elements are $R_{\sigma_i\rightarrow\sigma_j}$.
At $\tau\rightarrow \infty$ the composite system   eventually reaches to the steady state where
the probability distribution  does not alter with time and can be determined easily taking $\mathcal{R}\Pi_s({\sigma})=0$.  
 Now according to the construction of the  model, 
  those  steady state probability distributions  of two successive levels  are related by
  \begin{center}
   $\Pi_{s}(\sigma_1)= \exp^{-\beta_1 E_1}\Pi_{s}(\sigma_0) $,

    $\Pi_{s}(\sigma_2)= \exp^{-\beta_2 E_2}\Pi_{s}(\sigma_1) $,
    
        $\Pi_{s}(\sigma_3)= \exp^{-\beta_3 E_3}\Pi_{s}(\sigma_2) $,
      
  \end{center}
  and so on.   Therefore the steady state distribution for any state ($\sigma_i$) can be written as
 \begin{eqnarray}
     \Pi_{s}(\sigma_i)&&= \exp^{-\beta_i E_i}\Pi_{s}(\sigma_{i-1})\nn\\
    &&= \exp^{-(\beta_i E_i+.....+\beta_2 E_2 + \beta_1  E_1)}\Pi_{s}(\sigma_{0})\nn\\
    &&=\frac{e^{-\sum_{j=1}^i \beta_j E_j}}{Z}.
    \label{steady-den}
 \end{eqnarray}
 Where in the last line, we have used the normalization condition and  $Z=1+ \sum_{i=1}^M e^{-\sum_{j=1}^i \beta_j E_j}$.
  It is well known that as  time passes,  $\Pi_t$ (the probability density at any time t) will approach 
  monotonically towards $\Pi_s$ and their distance will reduce
 (Here for notational simplicity we have taken $0 \leq t < \tau$. However it will be true for any interval
 $(n-1)\tau \leq t < n\tau$). Hence one can write \cite{cover}

\begin{equation}
D(\Pi_{\tau}||\Pi_{s}) \leq D(\Pi_{0}||\Pi_{s}). 
\end{equation}
Here the Kullback-Leibler divergence is defined as 
$D(\Pi_{t_1}||\Pi_{t_2})=\sum_{i=0}^M \Pi_{t_1}(\sigma_i)\ln\frac{\Pi_{t_1}(\sigma_i)}{\Pi_{t_2}(\sigma_i)}$.
We can easily expand the above inequality and rewrite it to the form: 
\begin{equation}
 \sum_{i=0}^M [\Pi_{\tau}(\sigma_i)-\Pi_{0}(\sigma_i)] \ln \frac{1}{\Pi_{s}(\sigma_i)}
 \leq \mathcal{H}(\Pi_{\tau}) - \mathcal{H}(\Pi_{0}),
 \label{eq3}
\end{equation}
where $\mathcal{H}$ represents Shannon entropy which  have been already defined earlier. Therefore 
the right hand side represents change of Shannon entropy during the evolution. Again we have
\begin{equation}
 \ln \frac{1}{\Pi_{s}(\sigma_i)}=\ln Z + \sum_{j=1}^i \beta_j E_j.
\end{equation}
Then left hand side of eq.\ref{eq3} becomes
\begin{eqnarray}
 && \sum_{i=0}^M [\Pi_{\tau}(\sigma_i)-\Pi_{0}(\sigma_i)] \ln \frac{1}{\Pi_{s}(\sigma_i)}\nn\\
 &&=\sum_{i=1}^M [\Pi_{\tau}(\sigma_i)-\Pi_{0}(\sigma_i)] \sum_{j=1}^i \beta_j E_j\nn\\
 &&=\sum_{i=1}^M [\Pi_{\tau}(\sigma_i)-\Pi_{0}(\sigma_i)]  \beta_1 E_1
 +\sum_{i=2}^M [\Pi_{\tau}(\sigma_i)-\Pi_{0}(\sigma_i)]  \beta_2 E_2\nn\\
&& +.....+\sum_{i=j}^M [\Pi_{\tau}(\sigma_i)-\Pi_{0}(\sigma_i)]  \beta_j E_j
+....+ [\Pi_{\tau}(\sigma_M)-\Pi_{0}(\sigma_M)]  \beta_M E_M\nn\\.
\end{eqnarray}
Note that $\sum_{i=j}^M [\Pi_{\tau}(\sigma_i)-\Pi_{0}(\sigma_i)]$ represents the net change of 
probability of all the states above $(\sigma_j)$ including $(\sigma_j)$ during
the time of operation $0\leq t<\tau$. As all the allowed transitions form a linear chain, these net 
probability change can only happen if same amount of transitions occurs from $(\sigma_{j-1})$ state
to $(\sigma_j)$. Now for each of these transitions $E_j$ amount of heat will be absorbed from the 
bath $\beta_j$. Hence the average amount of heat that is absorbed from this bath along the evolution 
during time $\tau$ can be written as
\begin{equation}
 q_j=\sum_{i=j}^M [\Pi_{\tau}(\sigma_i)-\Pi_{0}(\sigma_i)] E_j.
\end{equation}
Therefore one can rewrite eq.\ref{eq3} as
\begin{equation}
 \beta_1 q_1 + \beta_2 q_2 +.....+\beta_M q_M\leq \mathcal{H}(\Pi_{\tau}) - \mathcal{H}(\Pi_{0}).
 \label{main-result}
\end{equation}
This is the second law for each individual cycle. The left side of the equation is related to the bath entropy production while
the right side represents entropy change of the joint system.
Now if the system is connected with a single bath and all the transitions are happening by 
exchanging energy with this bath, then  $\beta_1=\beta_2=...=\beta_M=\beta$ and 
 the above equation will simply reduce to 

\begin{equation}
 \beta q \le \mathcal{H}(\Pi_{\tau}) - \mathcal{H}(\Pi_{0}).
\end{equation}
where $q$ represents total heat absorbed from the bath. In \cite{mer16} it is assumed that each energy level
is associated to a work source, as a result,  the amount of heat absorbed in each transition from the single 
bath is equal to same amount of work extraction. Then the above result will be reduced to Eq.\ref{mer-res}
 as obtained in \cite{mer16}. Note that in \cite{mer16} transition between any two states is allowed.
However in our model, we have restricted it to accommodate multiple baths which act simultaneously
 on the system. To understand the applicability of our model, we  consider different examples in next section.
 But before going there, we will try to relate the right hand side of eq.{\ref{main-result}} with the
 entropy change of the tape for completeness of the paper. A detailed comparison had been provided in \cite{mer16}.
 Taking the notation of discrete process for $n^{th}$ cycle, eq.{\ref{main-result}}
 can be written as
 \begin{equation}
 \beta_1 q_1(n) + \beta_2 q_2(n) +.....+\beta_M q_M(n)\leq H(Y_n,S_{n+1}) - H(X_n,S_n).
 \end{equation}
 Here $q_i(n)$ represents heat absorbed from the $i^{th}$ bath in  $n^{th}$ cycle. 
 Now for N cycles, the above inequality becomes
 \begin{eqnarray}
 && \sum_{n=1}^N [ \beta_1 q_1(n) + \beta_2 q_2(n) +.....+\beta_M q_M(n)]\nn\\
 && \leq \sum_{n=1}^N [H(Y_n,S_{n+1}) - H(X_n,S_n)]\nn\\
 && = \sum_{n=1}^N [H(Y_n|S_{n+1}) - H(X_n|S_n)] + \sum_{n=1}^N [H(S_{n+1})-H(S_n))]\nn\\
 &&= \sum_{n=1}^N [H(Y_n|S_{n+1}) - H(X_n|S_n)] + H(S_{N+1})-H(S_1))\nn\\
  && \approx \sum_{n=1}^N [H(Y_n|S_{n+1})  - H(X_n|S_n)].
  \label{main-result2}
 \end{eqnarray}
 In the last line, it is assumed that  N is very large compared to the total number of joint states ($N\gg M$).
 As a result, the contribution of the system (demon) entropy (which will be order of $\ln M$) becomes
 negligible compared to the other terms. Note that,  the right hand side of the above equation is 
 not equal to the entropy change of the tape which is given by
 \begin{equation}
    H(Y^N)-  H(X^N)=\sum_{n=1}^{N} [H(Y_n|Y^{n-1})- H(X_n|X^{n-1})].
 \end{equation}
Hence our result differs from the earlier result \cite{boyd16,boyd17}, where the entropy production rate of the bath
is restricted by the change of these Shannon entropy rate (which includes all the correlation present in a stream of bits) 
between input tape and the processed output tape. 
Although  the correlation in the output tape may implicitly contain the information  how it is processed,
 the inequality in \cite{boyd16,boyd17}  had been derived  ignoring the detailed methodology for the  generation of the output bits. 
This concern has been  pointed out in  \cite{mer16} and consequently the second law has been derived. 
Moreover, it is shown that the obtained 
inequality is tight and can be approached  arbitrary close towards equality\cite{mer16}. 

Note that, in  \cite{bar13,bar14,man12,man13,ran16} the performance of the autonomous information machine 
had been described only taking the configuration entropy change of the tape; ignoring the correlation among the bits or the
possible correlation between the output tape and the demon that might be generated  during the operation. 
Now, it is generally assumed that the input sequence of the tape is not correlated with the demon state
i.e, $P(x_n|s_n)=P(x_n)$ then the right hand side of eq.(\ref{main-result2}) becomes
\begin{equation}
 \sum_{n=1}^N [H(Y_n|S_{n+1})  - H(X_n|S_n)]=\sum_{n=1}^N [H(Y_n|S_{n+1})  - H(X_n)].
\end{equation}
For simplicity we take uncorrelated input sequence and try to find out the differences between our result with 
that of \cite{bar13,bar14,man12,man13,ran16}.

 \subsection{uncorrelated input sequence}
 If the input sequence does not have any correlation, then $P(x^N)=P(x_N).....P(x_3)P(x_2)P(x_1)$ and the
 total entropy of the input tape now becomes
\begin{equation}
 H(x^N)=\sum_{n=1}^{N} H(X_n)=N H(X).
\end{equation}
 where $H(X_n)=-\sum_{x_n}P(x_n)\ln P(x_n)$. In the last step, it is assumed that the individual probability of 
 each element in a particular  position of the sequence (say $n^{th}$)  is independent of its position. 
 Again, eq.(\ref{main-result2}) can be rewritten in the form:
 \begin{eqnarray}
  &&\sum_{n=1}^N [ \beta_1 q_1(n) + \beta_2 q_2(n) +.....+\beta_M q_M(n)]\nn\\
   && \leq \sum_{n=1}^N [H(Y_n|S_{n+1}) - H(X_n|S_n)] + H(S_{N+1})-H(S_1))\nn\\
 &&  =  \sum_{n=1}^N [H(Y_n) - H(X_n)] - \sum_{n=1}^N I(S_{n+1},Y_n)+ H(S_{N+1})-H(S_1)).
 \end{eqnarray}
In last line it is again assumed that the input sequence  is uncorrelated with the demon states.
$I(S_{n+1},Y_n)$ represents the correlation between $S_{n+1}$ and $Y_{n}$ and is given by 
\begin{equation}
 I(S_{n+1},Y_n)=\sum_{s_{n+1}}\sum_{ y_n} P(s_{n+1},y_n) \ln \left(\frac{P(s_{n+1},y_n) }{P(s_{n+1})P(y_n) }\right).
\end{equation}
Note that, $I(S_{n+1},Y_n)$ is always positive. Neglecting the contribution of demon state (which becomes zero for large N)
the above inequality shows that our bound is more compact compared to the earlier one  \cite{bar13,bar14,man12,man13,ran16}.
Although in \cite{bar13,bar14,man12} only single heat bath has been taken, we are comparing the other part except the bath entropy.
Note that in \cite{man13} the author mentioned about the mutual information between the demon and tape but the exact expression has not been
given.

Now if the demon performs in steady state,
then there is no need to concern about each individual cycle (the average heat absorbed from $i^{th}$ bath
in $n^{th}$ cycle  $q_i(n)$ will be independent of the cycle i.e, $q_i(n)=q_i$). On the other hand entropy change of 
demon  will also be zero. Then the second law for uncorrelated independent sequence  becomes

\begin{equation}
  \beta_1 q_1 + \beta_2 q_2 +.....+\beta_M q_M \leq H(Y) - H(X)- \frac{1}{N}\sum_{n=1}^N I(S_{n+1},Y_n)).
\end{equation}.

For large $\tau$, the joint system may reach to the  steady state where probability distributions will take the 
form as shown in eq.\ref{steady-den}. Now if
the energy of each joint state can be written as sum of demon state energy and tape state energy, then 
 the corresponding total probability density can be expressed in terms of product of demon state probability and 
tape state probability. For this case, the correlation after the evolution at $\tau$ between 
demon state and tape state $I(S_{n+1},Y_n)$ vanishes.  Hence  the  second law in steady state for
uncorrelated independent sequence   in large  $\tau$ limit takes the form:
\begin{equation}
  \beta_1 q_1 + \beta_2 q_2 +.....+\beta_M q_M \leq H(Y) - H(X).
\end{equation}

In next section we will talk about few examples where our law can be applicable.

\section{Examples}

\subsection{Example 1}
 First we consider the Maxwell refrigerator model \cite{man13}.
 In this model,  a two level system is coupled with an information bath and two thermal baths.
 A simple binary tape is taken as an information reservoir. Hence depending on the system(demon) state and
 the bit state there will be four joint states 0d, 1d, 0u and 1u. Each bit can interact with the 
 demon for a time $\tau$ before the next bit arrives. The incoming bit can change
 its state only when it is interacting with the demon. After the interaction, for a time $\tau$, the bit
 retains its last state as an output and the tape is forwarded.
The rule for the transitions during the interaction time is as follows: When transition takes place with the 
exchange of heat with hot bath $T_h$, the bit state does not change, i.e., transition
between 0u and 0d, similarly between 1u and 1d energy is exchanged with bath $T_h$.
But for 0d and 1u energy is exchanged with the bath $T_c$. No other transition is permitted. 

Hence the joint states  form a linear chain and they are connected by the allowed transition:  
 $0u \leftrightarrows 0d \leftrightarrows 1u \leftrightarrows 1d$. Therefore  we can apply our model for this case.
Note that, in this model 0d state can exchange energy with bath $T_h$
in one end and it can also exchange energy with bath $T_c$ in another end.
In $\tau\rightarrow\infty$ corresponding steady state  density is given by
 
\begin{center}
 
$P_{s}(1u)=exp^{-E/T_h}P_{s}(1d)$,

$P_{s}(0d)=exp^{E/T_c}P_{s}(1u)$,

$P_{s}(0u)=exp^{-E/T_h}P_{s}(0d).$
 
\end{center}

Here, the second law for each cycle becomes $\beta_1 q_1 + \beta_2 q_2+\beta_3 q_3 \leq \mathcal{H}(\Pi_{\tau}) - \mathcal{H}(\Pi_{0})$.
Note that for this case, first and third bath is same i.e., $\beta_1=\beta_3=\frac{1}{k_B T_h}$ 
and second bath 
is denoted by $\beta_2=\frac{1}{k_B T_c}$. Then heat absorbed from hot and cold  bath is given by $q_h=q_1+q_3$ and 
$q_c=q_2$ respectively. When the incoming bits are uncorrelated, then, in  the steady state  the   second law simply reduces to 

\begin{equation}
 \frac{q_h}{T_h}+\frac{q_c}{T_c}\leq k_B\left[H(Y) - H(X) - \frac{1}{N}\sum_{n=1}^N I(S_{n+1},Y_n)\right].
\end{equation}

\subsection{Example 2}

In next autonomous  information machine model \cite{ran16}, there is an additional work source along with 
the information reservoir and two heat baths. 
The demon(system) consists of three states A, B and C.
 Hence depending on bit state and demon(system) state there will be six joint states.
 For any transition between the energy levels  A and B,  $E_1$ amount of energy is exchanged with bath $T_c$.
 This is true for any transition between B and C. Note that during these transitions bit 
 state is not changed. However, transition between A and C is restricted and depends on the interacting bit. 
  When  transition occurs from (to) C0 to (from) A1, E amount of energy is  absorbed (released)  from (to) bath $T_h$
  and $w$ amount of work is done on (extracted from) the system. 
  But Transition between C1 and A0 is restricted. Hence the allowed transitions, from one state to another, form a linear 
  chain and is given by $A0 \leftrightarrows B0 \leftrightarrows C0 \leftrightarrows A1 \leftrightarrows B1 \leftrightarrows C1$. 
  Therefore we can apply eq.\ref{main-result} also for this case.
  Similar to the earlier example, all the heat exchanged with the bath $T_c$  can be summed up to $q_c$ and heat 
  absorbed during the transition C0 and A1 is taken as $q_h$. Then the second  law takes the form:
 
 \begin{equation}
 \frac{q_h}{T_h}+\frac{q_c}{T_c}\leq k_B\left[H(Y) - H(X)  -\frac{1}{N} \sum_{n=1}^N I(S_{n+1},Y_n)\right].
\end{equation}
As average energy of the demon does not change here, hence the first law becomes $q_h+q_c=W$, where $W$ represents 
work extraction. On the other hand, there is no work source  in the earlier example and first law takes the 
form $q_h+q_c=0$, although second law is same. 

 \subsection{Example 3}
 Now If we take $E_1=0$, then the above problem will be reduced to the Maxwell demon model \cite{man12}.
 For any transition from A to B or B to C and vice versa, no energy is exchanged. Therefore,
  bath $T_c$ does not have any significance. Then the above second law will be reduced to
 
 \begin{equation}
 q_h \leq  k_B T_h\left[H(Y) - H(X)  - \frac{1}{N}\sum_{n=1}^N I(S_{n+1},Y_n)\right].
\end{equation}
 As total heat absorption $q_h$ will be equal to work extraction W then
 
 \begin{equation}
 W \leq k_B T_h\left[H(Y) - H(X)  - \frac{1}{N}\sum_{n=1}^N I(S_{n+1},Y_n)\right].
\end{equation}
Note that $k_B[H(Y) - H(X)]$ denotes the entropy change of the tape. As $I$ is always positive, 
maximum extractable work becomes less than that was previously thought\cite{man12} while writing same 
amount of information on tape. On the other hand, erasing same amount of information we need 
to do more work to compensate the term $I$.

\section{Numerical Results}

In this section we will prove our results numerically by considering the second example. 
 Lets define the weight parameter
 \begin{equation}
  \mathcal{E} = \tanh\left(\frac{E}{2 T_h}\right).
 \end{equation}
Note that $-1\le \mathcal{E}\le 1$.
Consider $\delta$ represents excess  of 0 in the input tape 
compared to the 1, i.e., 
\begin{equation}
 \delta=p(0)-p(1).
\end{equation}
Here $p(0)$ and $p(1)$ denotes the probability of $0$ and $1$ respectively in the incoming bit stream.
Note that $-1\le\delta\le1$. As the consecutive bits are uncorrelated with each other,
the Shannon entropy of the incoming tape becomes
\begin{equation}
 H(X)=-p(0)\ln p(0) -p(1)\ln p(1).
\end{equation}
which denotes information content per bit. Similarly,  if the probability to get $0$ and $1$ in the outgoing bit stream
are represented by $p'(0)$ and $p'(1)$; then the corresponding Shannon entropy will be 
\begin{equation}
 H(Y)=-p'(0)\ln p'(0) -p'(1)\ln p'(1).
\end{equation}
Then entropy change of the tape is given by
\begin{equation}
 \Delta S= k_B (H(Y) - H(X)).
\end{equation}
As $q_h$ and $q_c$ represents average heat absorption to the hot bath $T_h$ and cold bath $T_c$ 
per unit cycle in steady state, then  bath entropy production  will be 
\begin{equation}
 S_B=-\frac{q_h}{T_h} - \frac{q_c}{T_c}.
 \end{equation}
The hidden entropy generated due to the correlation of output bits and the demon per unit cycle in 
steady state is given by
\begin{equation}
 \Delta S_{cor}=- \frac{k_B}{N}\sum_{n=1}^N I(S_{n+1},Y_n)=-k_BI(S_{n+1},Y_n).
\end{equation}
Hence the second law for this case can be rewritten as 
\begin{equation}
 S_{tot}=S_B+  \Delta S +\Delta S_{cor}\ge 0.
\end{equation}
\begin{figure}[!ht]
\vspace{0.5cm}
\begin{center}
 \includegraphics[width=10cm]{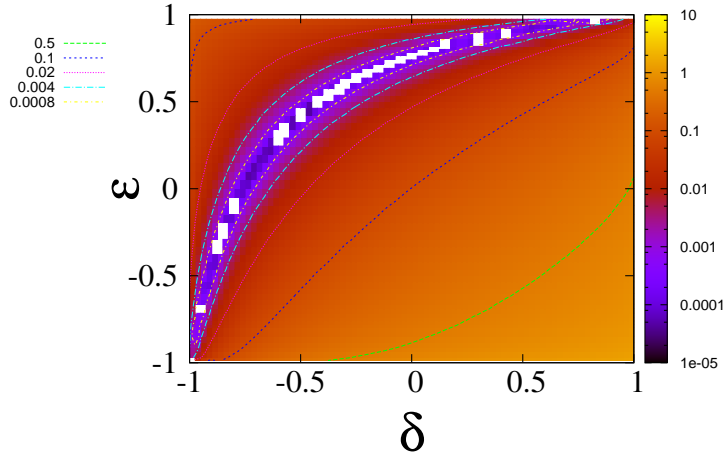}
\caption{ Histogram of $ S_{tot}$ for different values of $\delta$ and $\mathcal{E}$
at $T_h$  =  1.0, $T_c=0.5$ , $E_1$  =  0.5, $\tau =0.4$.}
\label{totent}
\end{center}
\end{figure}

\begin{figure}[!ht]
\vspace{0.5cm}
\begin{center}
 \includegraphics[width=10cm]{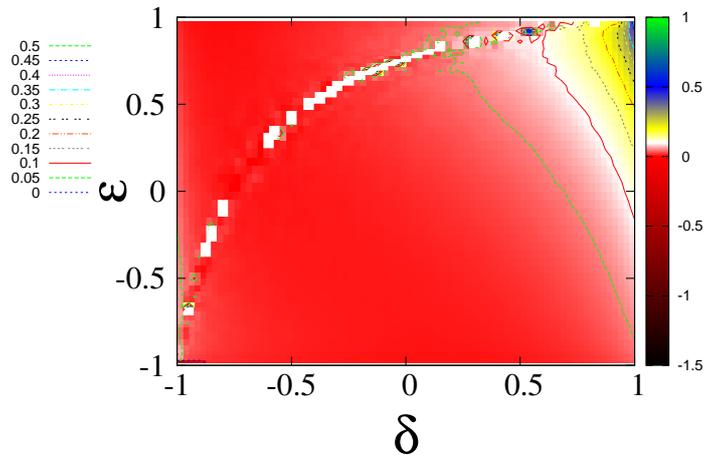}
\caption{   Histogram of $R_{err}$ for different values of $\delta$ and $\mathcal{E}$
at $T_h$  =  1.0, $T_c=0.5$, $E_1$ =0.5, $\tau=0.4$. }
\label{error}
\end{center}
\end{figure}

In numerical simulation we have set $T_h=1.0$, $T_c=0.5$ and $E_1=0.5$. We have also set 
Boltzmann constant $k_B=1$ and the interaction time of the each bit with the demon $\tau=0.4$.
In fig.\ref{totent} we have plotted total entropy production $S_{tot}$ for different parameter 
set $\delta$ and $\mathcal{E}$. We have found that it  remain always positive. Hence, it proves 
the second law. Moreover the figure clearly indicates the  reversible region, where 
$S_{tot}\rightarrow 0$, by simply connecting the white dots.

 The relative error  $R_{err}$ between the
the apparent entropy production $S_{app}=S_B+  \Delta S$ and $ S_{tot}$ is defined as
\begin{equation}
 R_{err}=\frac{S_{app}-S_{tot}}{S_{tot}}.
\end{equation}
 When the joint system behaves reversibly entropy production is 
zero and $R_{err}$ becomes undefined. The imaginary line connecting white dots in the fig.\ref{error}
denotes where these phenomena is occurring. Form the figure we have found that for quite large
region $R_{err}$ can take value greater than $10\%$ even it can exceed $50\%$ in certain region
($\delta\sim1$ and high $\mathcal{E}$). Hence we can not neglect  $S_{cor}$ term and $S_{tot}$ represents
the proper bound.

\section{Conclusions}
In summary we have studied an  autonomous Information engine model connecting with multiple baths, work 
source and information reservoir. This  is the most general scenario and second law has been obtained 
consequently. First we have derived the second law for each individual cycle. Then we sum those inequalities
to get the net inequality. The derivation is done taking into account the exact mechanism how an autonomous information machine
evolves connecting to the information reservoir. We have found that this inequality
is tighter compared to the earlier results  \cite{bar13,bar14,man12,man13,ran16} which only includes  configuration
entropy change of the tape between its  output and input sequences. Besides there  is a significant differences 
between our result with  \cite{boyd16,boyd17} where second law has been  derived ignoring the detailed operation. 
Moreover we have shown several examples where this law can be applicable. Finally from numerical simulation, we found  that 
the correction term which depends on the correlation between output tape and final state of the demon in each cycle,
is quite significant compared with the total entropy production and hence it should not be neglected.

\vspace{1cm}
{\large \bf Acknowledgements}
\normalsize

\vspace{0.5cm}
SR thanks A M Jayannavar for broad discussions throughout the work. SR also thanks  Deepak Dhar for useful discussions.
SR thanks Department of Science and Technology (DST), India for the financial support through SERB NPDF.

\end{document}